\documentclass[%
 reprint,
superscriptaddress,
nofootinbib,
 amsmath,amssymb,
 aps,
]{revtex4-1}

\usepackage{graphicx}
\usepackage{dcolumn}
\usepackage{bm}

\usepackage{ulem}
\usepackage{xcolor}

\usepackage[utf8]{inputenc}
\usepackage{graphicx}
\usepackage{epstopdf}
\usepackage{hyperref}
\usepackage{multirow}
\usepackage{geometry}
\usepackage{amsmath}
\allowdisplaybreaks[1]

\begin{document}

\begin{flushright}
LU TP 17-40 \\
\end{flushright}

\title{Analytic representation of $F_K/F_\pi$ in two loop chiral perturbation theory}

\author{B. Ananthanarayan}
\affiliation{Centre for High Energy Physics, Indian Institute of Science, Bangalore-560012, Karnataka, India}
\author{Johan Bijnens}
\affiliation{Department of Astronomy and Theoretical Physics, Lund University, S\"olvegatan 14A, SE 223-62 Lund, Sweden}
\author{Samuel Friot}
\affiliation{Institut de Physique Nucléaire d’Orsay, Université Paris-Sud 11, IN2P3-CNRS, F-91405 Orsay Cedex, France}
\affiliation{Institut de Physique Nucléaire de Lyon, Université Lyon 1, IN2P3-CNRS, F-69622 Villeurbanne Cedex, France}
\author{Shayan Ghosh}
\affiliation{Centre for High Energy Physics, Indian Institute of Science, Bangalore-560012, Karnataka, India}

\begin{abstract}
We present an analytic representation of $F_K/F_\pi$ as calculated in three-flavour two-loop chiral perturbation theory, which involves expressing three mass scale sunsets in terms of Kamp\'e de F\'eriet series. We demonstrate how approximations may be made to obtain relatively compact analytic representations. An illustrative set of fits using lattice data is also presented, which shows good agreement with existing fits.
\end{abstract}

\maketitle

{\bf Introduction}- The spectrum of QCD contains as lightest particles the pseudo-scalar
octet, and their properties provide a delicate test of its non-perturbative
features, including that of chiral symmetry breaking in the sector involving
the three lightest quarks. Of these, a special place is accorded to the decay
constants of the kaon and pion, namely $F_K$ and $F_\pi$. Their ratio has been
investigated on the lattice now, even at quark masses that include the physical values \cite{Durr:2016ulb}. On the other hand, in chiral perturbation theory (ChPT) \cite{Gasser:1984gg} at two-loops, expressions have been available for nearly two decades, but involving certain integrals (sunsets) that are evaluated numerically \cite{Amoros:1999dp}. In this work, we provide an analytic expression for $F_K/F_\pi$, which among other things incorporates double series derived using Mellin-Barnes (MB) representations of the sunsets. This allows us to produce a template for easy fitting to lattice simulations.

\vspace*{4mm}

{\bf Methodology}- Three-flavour ChPT expressions for the decay constants of the pseudoscalar
mesons at two-loops are given in \cite{Amoros:1999dp}. These may be decomposed
as:
\begin{align}
	\frac{F_P}{F_0} = 1 + F_P^{(4)} + \left( F_P \right)^{(6)}_{CT} + \left( F_P \right)^{(6)}_{loop} + \mathcal{O}(p^8) , \label{Eq:FP}
\end{align}
where $P$ is the particle in question. The $\mathcal{O}(p^6)$ contribution
can be subdivided as:
\begin{align}
	F_{\pi}^4 \left( F_P \right)^{(6)}_{loop} =&\, d_{sunset}^{P} + d_{log \times log}^{P} + d_{log}^{P} + d_{log \times L_i}^{P} \nonumber \\
	& + d_{L_i}^{P} + d_{L_i \times L_j}^{P} . \label{Eq:FPloop}
\end{align}
$d_{L_i\times log}^{P}$ collects the terms linear in the $\mathcal{O}(p^4)$
LECs $L_i$ and containing chiral logs,
$d_{log}^{P}$, $d_{log \times log}^{P}$ collect the terms
linear respectively quadratic in chiral logarithms without $L_i$,
$d_{L_i}$ and $d_{L_i \times L_j}^{P}$ the terms linear respectively quadratic
in the LECs $L_i$. The term $\left( F_P \right)^{(6)}_{CT}$ is composed of
the $\mathcal{O}(p^6)$ counterterms, i.e. the LECs $C^r_i$, while
$d_{sunset}^{P}$ are the pure sunset terms.

One determines the ratio $F_K/F_\pi$ using:
\begin{align}
	\frac{F_K}{F_{\pi}} &= 1 + \left( \frac{F_K}{F_0} \bigg|_{p^4} - \frac{F_{\pi}}{F_0} \bigg|_{p^4} \right)_{\text{NLO}} \nonumber \\
	& + \left( \frac{F_K}{F_0} \bigg|_{p^6} - \frac{F_{\pi}}{F_0} \bigg|_{p^6} - \frac{F_K}{F_0} \bigg|_{p^4} \frac{F_{\pi}}{F_0} \bigg|_{p^4} + \frac{F_{\pi}}{F_0} \bigg|^2_{p^4} \right)_{\text{NNLO}} . \label{Eq:fkfp}
\end{align}

The terms $d_{sunset}^{P}$ are not available fully analytically. Their determination is the goal of this work. The sunset integral is defined as:
\begin{align}
& {H}_{\{\alpha,\beta,\gamma\}}^d (m_1^2,m_2^2,m_3^2;p^2) =  \nonumber \\
& \frac{(1/i)^2}{(2\pi)^{2d}} \int \frac{d^dq \; d^dr}{[q^2-m_1^2]^{\alpha} [r^2-m_2^2]^{\beta} [(q+r-p)^2-m_3^2]^{\gamma}} .
\label{Eq:SunsetDef}
\end{align}
Aside from the basic scalar integral defined above, tensor integrals in which
the momenta $q_{\mu}$ and $q_{\mu} q_{\nu}$ appear in the numerator, and
derivatives with respect to the external momentum of both the scalar and tensor
integrals contribute to $d_{sunset}^{P}$ \cite{Amoros:1999dp}. The tensor integrals,
as well as all the derivatives, may be reduced into a linear combination of
scalar integrals using the methods given in \cite{Tarasov:1997kx}. Thus
only a smaller set of master integrals (MI) is needed.

The full list of sunset integrals contributing to $d_{sunset}^{P}$ can thus all be
expressed in terms of a set of four MI (${H}_{\{1,1,1\}}^d$, ${H}_{\{2,1,1\}}^d$,
${H}_{\{1,2,1\}}^d$ and ${H}_{\{1,1,2\}}^d$) and the one-loop tadpole integral. The
problem reduces to solving these analytically in the required mass
configurations. For the evaluation of $F_K/F_\pi$, seven distinct three mass
scale MI need evaluation.

MB theory leads to representations of these MI where each integral consists of
at least one double complex plane integral. These double MB integrals are
evaluated using the method proposed in \cite{Aguilar:2008qj} and fully
systematized in \cite{Friot:2011ic} to obtain results in the form of sums of
single and double infinite series \cite{Ananthanarayan:2016pos}-\cite{ABFG:2018}.

\vspace*{4mm}

{\bf The analytic representation}- Using Eq.(\ref{Eq:fkfp}), we obtain the following representation of $F_K/F_\pi$:
\begin{align}
\frac{F_K}{F_\pi} &= 1 + 4 (4 \pi )^2 L^r_5 \left(\xi _K-\xi _{\pi }\right)
 + \frac{5}{8} \xi_\pi \lambda_\pi - \frac{1}{4} \xi_K \lambda_K 
\nonumber \\
& + \left(\frac{1}{8} \xi_\pi - \frac{1}{2} \xi_K \right) \lambda_\eta
 + \xi_K^2 F_F\left[ \frac{m_\pi^2}{m_K^2} \right] + \hat K_1^r \lambda_\pi^2 
\nonumber \\
& + \hat K_2^r \lambda_\pi\lambda_K + \hat K_3^r \lambda_\pi\lambda_\eta
 + \hat K_4^r \lambda_K^2 + \hat K_5^r \lambda_K\lambda_\eta 
\nonumber \\
& + \hat K_6^r \lambda_\eta^2 \xi_K^2 + \hat C_1 \lambda_\pi 
+ \hat C_2 \lambda_K + \hat C_3 \lambda_\eta + \hat C_4 , \label{Eq:fkfpLattice}
\end{align}
where $\xi_\pi=m_\pi^2/(16\pi^2 F_\pi^2)$, $\xi_K= m_K^2/(16\pi^2 F_\pi^2)$,
$\lambda_i = \log(m_i^2/\mu^2)$, and:
\begin{align}
         \hat{K}^r_1 =\,& \frac{11}{24} \xi_\pi \xi_K - \frac{131}{192} \xi_\pi^2,
        &\hat{K}^r_2 =\,& -\frac{41}{96} \xi_\pi \xi_K - \frac{3}{32} \xi_\pi^2, \nonumber \\
         \hat{K}^r_3 =\,& \frac{13}{24} \xi_\pi \xi_K + \frac{59}{96} \xi_\pi^2 ,
        &\hat{K}^r_4 =\,& \frac{17}{36} \xi_K^2 + \frac{7}{144} \xi_\pi \xi_K,
 \nonumber \\
         \hat{K}^r_5 =\,& -\frac{163}{144} \xi_K^2 - \frac{67}{288} \xi_\pi \xi_K + \frac{3}{32} \xi_\pi^2 , \hspace*{-2cm} \nonumber \\
         \hat{K}^r_6 =\,& \frac{241}{288} \xi_K^2 - \frac{13}{72}  \xi_\pi \xi_K - \frac{61}{192} \xi_\pi^2 . \hspace*{-2cm}
\end{align}
\begin{align}
	& \hat{C}^r_1 = - \left(\frac{7}{9} + \frac{11}{2} (4 \pi )^2  L^r_{5} \right) \xi_\pi \xi_K\nonumber \\
	&  -\left(\frac{113}{72} + (4 \pi )^2 (4 L^r_{1} + 10 L^r_{2} + \frac{13}{2} L^r_{3} - \frac{21}{2} L^r_{5}) \right) \xi _\pi^2 , \nonumber \\[2mm]
	& \hat{C}^r_2 = \left(\frac{209}{144} + 3 (4\pi)^2 L^r_{5} \right) \xi_\pi \xi_K \nonumber \\
	&  + \left(\frac{53}{96} + (4 \pi )^2 (4 L^r_{1} + 10 L^r_{2} + 5 L^r_{3} - 5 L^r_{5}) \right) \xi _K^2 , \nonumber \\[2mm]
	& \hat{C}^r_3 = \left( \frac{13}{18}  + (4 \pi )^2 \left( \frac{8}{3} L^r_{3} - \frac{2}{3} L^r_{5} - 16 L^r_{7} - 8 L^r_{8} \right) \right) \xi_K^2  \nonumber \\
	& - \left( \frac{4}{9} + (4\pi)^2 \left( \frac{4}{3} L^r_{3} + \frac{25}{6} L^r_{5} - 32 L^r_{7} - 16 L^r_{8} \right) \right) \xi _\pi \xi_K \nonumber \\
	& + \left( \frac{19}{288} +  (4 \pi)^2 \left( \frac{1}{6} L^r_{3} + \frac{11}{6} L^r_{5} - 16 L^r_{7} - 8 L^r_{8} \right) \right) \xi_\pi^2 , \nonumber \\[2mm]
	& \hat{C}^r_4 = (4 \pi)^2 (\xi_K - \xi_\pi) \nonumber \\
	& \times \bigg\{ 8 (4 \pi )^2 \bigg( 2 (C^r_{14}+C^r_{15}) \xi _K +  (C^r_{15}+2 C^r_{17}) \xi_\pi \bigg) \nonumber \\
	& + \bigg( 8 (4 \pi )^2 L^r_{5} (8 L^r_{4}+3 L^r_{5}-16 L^r_{6}-8 L^r_{8})- 2 L^r_{1} \nonumber \\
	& \quad - L^r_{2} - \frac{1}{18} L^r_{3} + \frac{4}{3} L^r_{5} - 16 L^r_{7} - 8 L^r_{8}  \bigg) \xi_K \nonumber \\
	& + \bigg( 8 (4 \pi )^2 L^r_{5} (4 L^r_{4} + 5 L^r_{5} - 8 L^r_{6} - 8 L^r_{8}) - 2 L^r_{1}  \nonumber \\
	& \quad - L^r_{2} - \frac{5}{18} L^r_{3} - \frac{4}{3} L^r_{5} + 16 L^r_{7} + 8 L^r_{8} \bigg) \xi _{\pi } \bigg\}.
\end{align}

$F_F$ consists of the terms arising from the pure sunset contributions. 
The split between the $\hat K_i$ terms and $F_F$ is not unique: one convenient decomposition, that takes into account the freedom to distribute the chiral logs while keeping the final result unchanged, is:
\begin{align}
	& F_F = \frac{m_\pi^6}{m_K^6} \left(\frac{49}{48}+\frac{\pi ^2}{32}\right) + \frac{m_\pi^4}{m_K^4} \left(\frac{25871}{6912}+\frac{919 \pi^2}{2592}\right) \nonumber \\
	& -\frac{m_\pi^2}{m_K^2} \left(\frac{9875}{864}+\frac{757 \pi ^2}{1296}\right) +  \left(\frac{39233}{6912}+\frac{437 \pi ^2}{1296}\right) \nonumber \\ 
	&  +\frac{m_K^2}{m_\pi^2} \left(\frac{3}{2}-\frac{\pi ^2}{12}\right) -\frac{3}{32} \log ^2\left[\frac{m_\pi^2}{m_K^2}\right] - \frac{9}{16} \log \left[\frac{m_\pi^2}{m_K^2}\right] \nonumber \\
	& - \frac{1}{8} \frac{m_K^2}{m_\pi^2} \log ^2\left[\frac{4}{3}-\frac{m_\pi^2}{3 m_K^2}\right] + \frac{5}{64} \frac{m_\pi^6}{m_K^6} \log ^2\left[\frac{4 m_K^2}{3 m_\pi^2}-\frac{1}{3}\right] \nonumber \\
	& + \frac{(16\pi^2)^2}{m_K^4} \left( d^K_{K \pi \pi} + d^K_{K \eta \eta} + d^K_{K \pi \eta} - d^\pi_{\pi K K} - d^\pi_{\pi \eta \eta} - d^\pi_{K K \eta} \right)
\label{Eq:ExactFf}
\end{align}
where:
\begin{align}
	d^K_{K \pi \pi} &=  -\left(\frac{27}{64} \frac{m_{\pi}^4}{m_{K}^2} + \frac{1}{64}m_{K}^2 + \frac{9}{16} m_{\pi}^2 \right) \overline{H}^K_{K \pi \pi} \nonumber \\
	& + \left(\frac{1}{16} m_{K}^4 + \frac{1}{8} m_{K}^2 m_{\pi}^2 + \frac{9}{16}  m_{\pi}^4 \right) \overline{H}^K_{2K \pi \pi},
\end{align}
\begin{align}
	d^K_{K \eta \eta} &= - \left( \frac{15}{64} \frac{m_{\pi}^4}{m_{K}^2} + \frac{1189}{576} m_{K}^2 - \frac{65}{48} m_{\pi}^2 \right) \overline{H}^K_{K \eta \eta} \nonumber \\
	& + \left(\frac{143}{48} m_{K}^4 - \frac{139}{72} m_{K}^2 m_{\pi}^2 + \frac{5}{16} m_{\pi}^4 \right) \overline{H}^K_{2K \eta \eta},
\end{align}
\begin{align}
	d^K_{K \pi \eta} &= \left( - \frac{7}{32} \frac{m_{\pi}^4}{m_{K}^2} + \frac{5}{96} m_{K}^2 + \frac{7}{6} m_{\pi}^2 \right) \overline{H}^{K}_{K \pi \eta} \nonumber \\
	& + \left( \frac{3}{8} \frac{m_{\pi}^6}{m_{K}^2} + \frac{1}{4} m_{K}^2 m_{\pi}^2 - \frac{15}{8} m_{\pi}^4 \right) \overline{H}^{K}_{K 2\pi \eta} \nonumber \\
	& - \left( \frac{11}{18} m_{K}^4 - \frac{1}{12} \frac{m_{\pi}^6}{m_{K}^2} + \frac{41}{72} m_{K}^2 m_{\pi}^2 + \frac{11}{72}m_{\pi}^4 \right) \overline{H}^{K}_{K \pi 2\eta} \nonumber \\
	&  - \left( \frac{1}{2} m_{K}^4 \right) \overline{H}^{K}_{2K \pi \eta},
\end{align}
\begin{align}
	{d}^{\pi}_{\pi K K} & = - \left(\frac{9}{16} \frac{m_{K}^4}{m_{\pi}^2} + \frac{3}{4} m_{K}^2 + \frac{1}{48} m_{\pi}^2 \right) \overline{H}^{\pi}_{\pi K K} \nonumber \\
	& + \left( \frac{3}{4} m_{K}^4 + \frac{1}{6} m_{K}^2 m_{\pi}^2 +\frac{1}{12} m_{\pi}^4 \right) \overline{H}^{\pi}_{2\pi K K},
\end{align}
\begin{align}
	{d}^{\pi}_{\pi \eta \eta} &= \left( -\frac{1}{36} m_{\pi}^2 \right) \overline{H}^{\pi}_{\pi \eta \eta}+\left( \frac{1}{36} m_{\pi}^4 \right) \overline{H}^{\pi}_{2\pi \eta \eta},
\end{align}
and
\begin{align}
	{d}^{\pi}_{K K \eta} &= \left( \frac{15}{16} \frac{m_{K}^4}{m_{\pi}^2} - \frac{13}{36} m_{K}^2 + \frac{13}{144} m_{\pi}^2 \right) \overline{H}^{\pi}_{K K \eta} \nonumber \\
	& + \left( \frac{91}{108} m_{K}^4 - \frac{m_{K}^6}{m_{\pi}^2} - \frac{5}{27} m_{K}^2 m_{\pi}^2 + \frac{m_{\pi}^4 }{108}\right) \overline{H}^{\pi}_{K K 2\eta} \nonumber \\	
	& + \left( \frac{1}{2} m_{K}^4 - 2 \frac{m_{K}^6}{m_{\pi}^2} - \frac{1}{6} m_{K}^2 m_{\pi}^2 \right) \overline{H}^{\pi}_{2K K \eta}.
\end{align}

\begin{figure}[tb]
\centering
\includegraphics[width=0.25\textwidth]{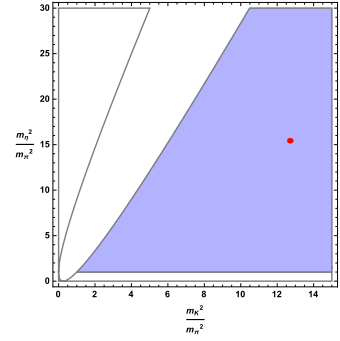}
\caption{Region of convergence of Eqs.(\ref{Eq:Hkpe})-(\ref{Eq:Hekk}) (blue region). The red dot marks the physical values of the meson masses.}
\label{Fig:Convergence}
\end{figure}

The MI are denoted by $\overline{H}^{S}_{aP \, bQ \, cR} \equiv \overline{H}^d_{\{a,b,c\}}(m_P^2,m_Q^2,m_R^2;p^2=m_S^2 )$, the ``bar" indicating that the chiral subtraction prefactor $\left( \mu^2 \frac{e^{\gamma_E-1}}{4\pi} \right)^{4-d}$ has been taken into acount and that the chiral logarithms have been extracted and included in the log terms of Eq.(\ref{Eq:FPloop}). Expressions for the two mass scale MI are given in \cite{Ananthanarayan:2017yhz}, and those for the three mass scale are given below in terms of generalized hypergeometric (${}_pF_q$) and Kamp\'e de F\'eriet (KdF) series. The three mass scale MI not explicitly presented here can be derived from the following by differentiation w.r.t the appropriate square propagator mass. The validity of Eqs.(\ref{Eq:Hkpe})-(\ref{Eq:Hekk}) is dictated by the region of convergence of the KdF and ${}_pF_q$ series, which is given by $(m_\pi<m_\eta) \wedge (m_\pi+m_\eta<2m_K)$ and shown in Fig.~\ref{Fig:Convergence}.

\begin{align}
\label{Eq:Hkpe}
	& \overline{H}^{K}_{K \pi \eta} = \frac{m_{K}^2}{512\pi ^4} \Bigg\{ - \frac{7}{4}\left(\frac{m_{\eta}^4}{m_{K}^4}+\frac{m_{\pi}^4}{m_{K}^4}\right)  -\frac{m_{\pi}^2}{m_{K}^2} \log\left[\frac{m_{\pi}^2}{m_{K}^2}\right]^2  \nonumber \\
	& + \left(1-\frac{\pi^2}{2}\right)\left(\frac{m_{\eta}^2}{m_{K}^2}+\frac{m_{\pi}^2}{m_{K}^2}\right) +\frac{m_{\pi}^4}{2 m_{K}^4} \log\left[\frac{m_{\pi}^2}{m_{K}^2}\right] -\frac{1}{4} \nonumber \\
	& +\frac{m_{\pi}^2}{m_{K}^2} \frac{m_{\eta}^2}{m_{K}^2} \bigg( 7+\frac{2 \pi^2}{3}-2 \log\left[\frac{m_{\eta}^2}{m_{K}^2}\right]-2 \log\left[\frac{m_{\pi}^2}{m_{K}^2}\right] \nonumber \\
	& +\log\left[\frac{m_{\eta}^2}{m_{K}^2}\right] \log\left[\frac{m_{\pi}^2}{m_{K}^2}\right] \bigg) + \frac{m_{\eta}^4}{2 m_{K}^4} \log\left[\frac{m_{\eta}^2}{m_{K}^2}\right] +  \frac{5 \pi^2}{6} \nonumber \\
	& -\frac{m_{\eta}^2}{m_{K}^2} \log\left[\frac{m_{\eta}^2}{m_{K}^2}\right]^2 +\frac{8 \pi }{3}\left(\frac{m_{\eta}^2}{m_{K}^2}\right)^{3/2} 
		{}_2F_1 \bigg[ \begin{array}{c}
		\frac{1}{2},-\frac{1}{2} \\
		\frac{5}{2} \\
	\end{array}	\bigg| \frac{m_\eta^2}{4m_K^2} \bigg] \nonumber \\
	& 
	+\frac{1}{36}\frac{m_{\eta}^6}{m_{K}^6}
		{}_3F_2 \bigg[ \begin{array}{c}
		1,1,2 \\
		\frac{5}{2},4 \\
	\end{array}	\bigg| \frac{m_\eta^2}{4m_K^2} \bigg]
	 + \frac{1}{36} \frac{m_{\pi}^6}{m_{K}^6}
		{}_3F_2 \bigg[ \begin{array}{c}
		1,1,2 \\
		\frac{5}{2},4 \\
	\end{array}	\bigg| \frac{m_\pi^2}{4m_K^2} \bigg]  \nonumber \\
	& + \frac{1}{6} \frac{m_{\eta}^4}{m_K^4} \frac{m_{\pi}^2}{m_K^2}
	 \left( 2\gamma_E - 1 + \log \left[\frac{m_{\pi}^2 m_{\eta}^2}{16 m_K^4}\right] \right) {}_2F_1 \bigg[ \begin{array}{c}
		1,1 \\
		\frac{5}{2} \\
	\end{array}	\bigg| \frac{m_\pi^2}{4m_K^2} \bigg] \nonumber \\
	& + \frac{\sqrt{\pi}}{8} \frac{m_{\pi}^2}{m_{K}^2} \frac{m_{\eta}^4}{m_{K}^4} \left( \log \left[\frac{m_{\eta}^2}{4 m_{K}^2}\right]+\log \left[\frac{m_{\pi}^2}{4 m_{K}^2}\right] + \frac{\partial}{\partial \alpha} \right) \cdot \nonumber\\
	& \Bigg( \frac{\Gamma(1+2\alpha) \Gamma(2+2\alpha) \Gamma(3+2\alpha)}{\Gamma(1+\alpha) \Gamma^2(2+\alpha) \Gamma(3+\alpha) \Gamma(\frac{5}{2}+2\alpha)} \nonumber\\
	& F^{3:1}_{1:2} \bigg[ \begin{array}{c}
		1+2\alpha, 2+2\alpha, 3+2\alpha: 1,1 \\
		\frac{5}{2}+2\alpha: 2+\alpha, 1+\alpha; 3+\alpha, 2+\alpha \\
	\end{array}	\bigg| \frac{m_\eta^2}{4m_K^2} , \frac{m_\pi^2}{4m_K^2} \bigg] \Bigg) \Bigg|_{\alpha=0} \nonumber \\
	& - \frac{m_K}{m_\eta} \frac{m_{\pi}^4}{m_{K}^4} \left( \log \left[\frac{m_{\pi}^2}{m_{\eta}^2} \right] + \frac{\partial}{\partial \alpha} \right) \cdot \Bigg( \frac{\Gamma(\frac{1}{2}+\alpha) \Gamma(\frac{3}{2}+\alpha)}{\Gamma(2+\alpha) \Gamma(3+\alpha)} \nonumber \\
	& F^{0:3}_{2:0} \bigg[ \begin{array}{c}
		- : \frac{1}{2}+\alpha,-\frac{1}{2}; \frac{3}{2}+\alpha,\frac{1}{2}; 1,\frac{3}{2} \\
		2+\alpha, 3+\alpha : - \\
	\end{array}	\bigg| \frac{m_\pi^2}{m_\eta^2}, \frac{m_\pi^2}{4m_K^2} \bigg] \Bigg) \Bigg|_{\alpha=0} \nonumber \\
	& + \frac{m_\pi^2}{m_K^2} \frac{m_\eta}{m_K} \left( \log \left[\frac{m_{\pi}^2}{m_{\eta}^2} \right] + \frac{\partial}{\partial \alpha} \right)  \cdot \nonumber\\
	& \Bigg( \frac{\pi^2 }{\Gamma(\frac{1}{2}-\alpha) \Gamma(\frac{3}{2}-\alpha) \Gamma(1+\alpha) \Gamma(2+\alpha)} \nonumber \\
	& F^{3:1}_{1:2} \bigg[ \begin{array}{c}
		-\frac{1}{2},\frac{1}{2},\frac{3}{2}:1,1 \\
		1: \frac{1}{2}-\alpha,1+\alpha; \frac{3}{2}-\alpha, 2+\alpha \\
	\end{array}	\bigg| \frac{m_\eta^2}{4m_K^2}, \frac{m_\pi^2}{4m_K^2} \bigg] \Bigg) \Bigg|_{\alpha=0} \nonumber \\
	& + \frac{\sqrt{\pi}}{16} \frac{m_\eta^2}{m_K^2} \frac{m_\pi^4}{m_K^4} \frac{\partial}{\partial \alpha} \cdot \Bigg( \frac{\Gamma(1+2\alpha) \Gamma(2+\alpha) \Gamma(3+\alpha)}{\Gamma(\frac{5}{2}+2\alpha)} \nonumber \\
	& \quad {}_4F_3
	\bigg[ \begin{array}{c}
		1, 1+2\alpha, 2+\alpha, 3+\alpha \\
		2,3,\frac{5}{2}+2\alpha \\
	\end{array}	\bigg| \frac{m_\pi^2}{4m_K^2} \bigg] \Bigg) \Bigg|_{\alpha=0} \Bigg\},
\end{align}
\begin{align}
	& \overline{H}^{K}_{2K \pi \eta} = \frac{1}{512\pi ^4} \Bigg\{ -\frac{m_{\eta}^2}{m_{K}^2} \bigg( 1 + \frac{\pi^2}{3} + \frac{1}{2} \log ^2 \left[\frac{m_{K}^2}{m_{\eta}^2}\right] \nonumber \\
	&  + \log \left[\frac{m_{K}^2}{m_{\eta}^2}\right] + \text{Li}_2 \left[ 1-\frac{m_{\pi}^2}{m_{\eta}^2} \right] \bigg) -\frac{m_{\pi}^2}{m_{K}^2} \bigg( 1 + \frac{\pi^2}{3} \nonumber \\
	& - \log \left[\frac{m_{\pi}^2}{m_{K}^2}\right] - \log \left[\frac{m_{K}^2}{m_{\eta}^2}\right] \log \left[\frac{m_{\pi}^2}{m_{K}^2}\right] - \frac{1}{2} \log^2 \left[ \frac{m_{K}^2}{m_{\eta}^2} \right]  \nonumber \\
	&  - \text{Li}_2 \left[1-\frac{m_{\pi}^2}{m_{\eta}^2}\right] \bigg) + \frac{2 \pi}{3} \frac{m_{\eta}^3}{m_{K}^3}
	{}_2F_1 \bigg[ \begin{array}{c}
		\frac{1}{2},\frac{1}{2} \\
		\frac{5}{2} \\
	\end{array}	\bigg| \frac{m_\eta^2}{4m_K^2} \bigg]  \nonumber \\
	& - \frac{m_{\pi}^4}{4 m_{K}^4}
	{}_3F_2 \bigg[ \begin{array}{c}
		1,1,1 \\
		\frac{3}{2},3 \\
	\end{array}	\bigg| \frac{m_\pi^2}{4m_K^2} \bigg]
	 - \frac{m_{\eta}^4}{4 m_{K}^4} 
	 {}_3F_2 \bigg[ \begin{array}{c}
		1,1,1 \\
		\frac{3}{2},3 \\
	\end{array}	\bigg| \frac{m_\eta^2}{4m_K^2} \bigg] \nonumber \\
	& -\frac{\sqrt{\pi}}{4} \frac{m_\eta^2}{m_K^2} \frac{m_\pi^2}{m_K^2} \left( \log \left[ \frac{m_\pi^2}{4m_K^2} \right] + \log \left[ \frac{m_\eta^2}{4m_K^2} \right] + \frac{\partial}{\partial \alpha} \right) \cdot \nonumber \\
	& \Bigg( \frac{\Gamma^2(1+2\alpha) \Gamma(2+2\alpha)}{\Gamma(\frac{3}{2}+2\alpha) \Gamma^2(1+\alpha) \Gamma^2(2+\alpha)} \nonumber \\
	& F^{3:1}_{1:2} \bigg[ \begin{array}{c}
		1+2\alpha,1+2\alpha,2+2\alpha:1,1 \\
		\frac{3}{2}+2\alpha: 1+\alpha,1+\alpha; 2+\alpha, 2+\alpha \\
	\end{array}	\bigg| \frac{m_\eta^2}{4m_K^2}, \frac{m_\pi^2}{4m_K^2} \bigg] \Bigg) \Bigg|_{\alpha=0}	\nonumber \\ 	
	& + \frac{5 \pi^2}{6} -1 + \frac{\pi^2}{4} \frac{m_\eta}{m_K} \frac{m_\pi^2}{m_K^2} \left( \log \left[ \frac{m_\pi^2}{m_\eta^2} \right] + \frac{\partial}{\partial \alpha} \right) \cdot \nonumber \\
	& \Bigg( \frac{1}{\Gamma(\frac{1}{2}-\alpha) \Gamma(\frac{3}{2}-\alpha) \Gamma(1+\alpha) \Gamma(2+\alpha)} \nonumber \\
	&  F^{3:1}_{1:2} \bigg[ \begin{array}{c}
		\frac{1}{2},\frac{1}{2},\frac{3}{2}:1,1 \\
		1: \frac{1}{2}-\alpha,1+\alpha; \frac{3}{2}-\alpha, 2+\alpha \\
	\end{array}	\bigg| \frac{m_\eta^2}{4m_K^2}, \frac{m_\pi^2}{4m_K^2} \bigg] \Bigg) \Bigg|_{\alpha=0}	\nonumber \\
	-& \frac{1}{4} \frac{m_\pi}{m_\eta} \frac{m_\pi^3}{m_K^3} \left( \log \left[ \frac{m_\pi^2}{m_\eta^2} \right] + \frac{\partial}{\partial \alpha} \right) \cdot \bigg( \frac{\Gamma(\frac{1}{2}+\alpha)\Gamma(\frac{3}{2}+\alpha)}{\Gamma(2+\alpha)\Gamma(3+\alpha)} \nonumber \\
	& F^{0:3}_{2:0} \bigg[ \begin{array}{c}
		-: 1,\frac{1}{2}; \frac{1}{2}+\alpha,\frac{1}{2}; \frac{3}{2}+\alpha, \frac{3}{2} \\
		2+\alpha, 3+\alpha: - \\
	\end{array}	\bigg| \frac{m_\pi^2}{m_\eta^2}, \frac{m_\pi^2}{4m_K^2} \bigg] \bigg) \Bigg|_{\alpha=0} \Bigg\},
\label{Eq:H2kpe}
\end{align}
and
\begin{align}
	& \overline{H}^{\pi}_{K K \eta} = \frac{m_\eta^2}{512 \pi ^4} \Bigg\{ \frac{\pi ^2}{6}-5 + 4 \log \left[\frac{m_\eta^2}{m_K^2}\right] - \log ^2 \left[\frac{m_\eta^2}{m_K^2}\right] \nonumber \\
	&  + \frac{m_K^2}{m_\eta^2} \left(6 + \frac{\pi ^2}{3}\right) - \frac{1}{18} \frac{m_\pi^2}{m_K^2} \frac{m_\pi^2}{m_\eta^2} {}_3F_2 \bigg[ \begin{array}{c}
		1,1,2 \\
		\frac{5}{2},4 \\
	\end{array}	\bigg| \frac{m_\pi^2}{4m_K^2} \bigg]  \nonumber \\
	& + \frac{m_\pi^2}{m_\eta^2} \left( \log \left[ \frac{m_K^2}{m_\pi^2} \right] + \frac{5}{4} \right) - \frac{\sqrt{\pi}}{8} \left( \log \left[ \frac{m_\eta^2}{4 m_K^2} \right] + \frac{\partial}{\partial \alpha} \right)\cdot  \nonumber \\
	& \Bigg(
	\frac{m_\pi^2}{m_K^2} \frac{\Gamma(3+\alpha)}{\Gamma(\frac{5}{2}+\alpha)} F^{3:1}_{1:2} \bigg[ \begin{array}{c}
		1+\alpha,2+\alpha,3+\alpha:1,1 \\
		\frac{5}{2}+\alpha:2,1+\alpha;3,2+\alpha \\
	\end{array}	\bigg| \frac{m_\pi^2}{4m_K^2}, \frac{m_\eta^2}{4m_K^2} \bigg] \nonumber \\
	& + \frac{2 m_\eta^2}{m_K^2} \frac{\Gamma(1+\alpha)}{\Gamma(\frac{5}{2}+\alpha)} {}_2F_1 \bigg[ \begin{array}{c}
		1,1+\alpha \\
		\frac{5}{2}+\alpha \\
	\end{array}	\bigg| \frac{m_\eta^2}{4m_K^2} \bigg] \Bigg) \Bigg|_{\alpha=0}  \Bigg\}.
\label{Eq:Hekk}
\end{align}

One may obtain simplified representations for $F_F$ by truncating the series at the desired precision, and taking an expansion around $ \rho= \frac{m_\pi^2}{m_K^2} = 0$. For illustrative purposes, we present one such representation in which we truncate the series such that the error between the exact and truncated values is $<1\%$ for most of the sets of masses used in the lattice study of \cite{Durr:2016ulb}. We get:
\begin{align}
	F_F & ( \rho ) =  a_1 + \left( a_2 + a_3 \log[\rho] + a_4 \log^2[\rho] \right) \rho \nonumber \\
	&  + \left( a_5 + a_6 \log[\rho] + a_7 \log^2[\rho] \right) \rho^2 \nonumber\\
	&  + \left( a_8 + a_9 \log[\rho] + a_{10} \log^2[\rho] \right) \rho^3 \nonumber \\
	&  + \left( a_{11} + a_{12} \log[\rho] + a_{13} \log^2[\rho] \right) \rho^4  + \mathcal{O} \left( \rho^5 \right) \label{Eq:ApproxFF}
\end{align}
where:
\begin{align}
	a_1 &= -\frac{6337}{5184} \left(\text{Li}_2\left[ \frac{3}{4} \right]+\log (4) \log \left[\frac{4}{3}\right] \right) + \frac{41 \pi ^2}{192} \nonumber \\
	& -\frac{11 \sqrt{2} \pi }{27} +\frac{85957107031}{27662342400}-\frac{119 \pi }{216 \sqrt{2}}   \nonumber \\
	& +\frac{62591}{612360}  \log [3] +\frac{43006343}{13471920} \log \left[\frac{4}{3}\right]  \nonumber \\
	& +\left(\frac{8 \sqrt{2}}{9}-\frac{41 \pi }{48}-\frac{5 \log [3]}{24 \sqrt{2}}\right) \csc ^{-1}\left[\sqrt{3}\right] \nonumber \\
	& +\frac{41}{48} \csc ^{-1}\left[\sqrt{3}\right]^2 + \frac{5}{1152} \log ^2 \left[\frac{4}{3}\right] , \nonumber \\[2mm]
	a_2 &= \frac{5821}{2592} \left(\text{Li}_2\left[\frac{3}{4}\right]+\log [4] \log \left[\frac{4}{3}\right]\right) - \frac{25 \pi ^2}{96} \nonumber \\
	& -\frac{7269419973251}{1120324867200}+\frac{145 \pi }{72 \sqrt{2}}+\frac{38693 \pi }{25920 \sqrt{3}}+\frac{82 \gamma }{405} \nonumber \\
	& -\frac{121}{576} \log ^2\left[\frac{4}{3}\right]-\left(\frac{6035437}{9797760}+\frac{13 \pi }{864 \sqrt{3}}\right) \log [3] \nonumber \\
	& -\left(\frac{468002719}{161663040}+\frac{13 \pi }{576 \sqrt{3}}\right) \log \left[\frac{4}{3}\right] -\frac{29}{324} \psi\left[\frac{5}{2}\right] \nonumber \\
	& + \left(\frac{463 \log [3]}{384 \sqrt{2}} + \frac{\log \left[\frac{4}{3}\right]}{2 \sqrt{2}} - \frac{11 \pi }{48}-\frac{13 \gamma }{18 \sqrt{2}}-\frac{15875}{3456 \sqrt{2}}\right) \nonumber \\
	& \quad \times \csc ^{-1}\left[\sqrt{3}\right] + \frac{11}{48} \csc ^{-1}\left[\sqrt{3}\right]^2 , \nonumber \\[2mm]
	a_3 &= \frac{803}{810}+\frac{13 \pi }{1728 \sqrt{3}}+\frac{7}{48} \log \left[\frac{4}{3}\right] - \frac{1}{2 \sqrt{2}} \csc ^{-1}\left[\sqrt{3}\right] , \nonumber \\[2mm]
	a_4 &= -\frac{11}{24} , \quad a_7 = \frac{337}{384} , \quad a_{10} = -\frac{9}{64} , \quad a_{13} = -\frac{27}{128}
	\nonumber \\[2mm]
	a_5 &= \frac{47}{128} \log ^2\left[\frac{4}{3}\right] -\frac{845}{648} \left(\text{Li}_2\left[\frac{3}{4}\right]+\log [4] \log \left[\frac{4}{3}\right] \right)  \nonumber \\
	& -\frac{1301 \sqrt{3} \pi }{512}-\frac{66191 \gamma }{12960}+\frac{1576413731881}{3585039575040}  + \frac{5 \pi ^2}{18} \nonumber \\
	& -\frac{145 \pi }{144 \sqrt{2}}+\frac{3572063 \pi }{663552 \sqrt{3}} + \frac{59}{48} \csc ^{-1}\left[\sqrt{3}\right]^2   \nonumber \\
	& + \left(\frac{744674317}{313528320}+\frac{176189 \pi }{55296 \sqrt{3}}\right) \log [3] +\frac{35}{144} \psi \left[\frac{5}{2}\right]  \nonumber \\
	& + \bigg(\frac{97621}{55296 \sqrt{2}} -\frac{59 \pi }{48} + \frac{3167 \gamma }{288 \sqrt{2}}-\frac{19589 \log [3]}{4096 \sqrt{2}} \nonumber \\
	&  \quad -\frac{115}{48 \sqrt{2}} \bigg) \log \left[\frac{4}{3}\right] \csc^{-1} \left[\sqrt{3}\right]  \nonumber \\
	& + \left(\frac{4312709021}{1293304320}+\frac{176189 \pi }{36864 \sqrt{3}}\right) \log \left[\frac{4}{3}\right] , \nonumber \\[2mm]
	a_6 &= \frac{17003}{8640}-\frac{176189 \pi }{110592 \sqrt{3}}-\frac{155}{192} \log \left[\frac{4}{3}\right] \nonumber \\ & + \frac{115}{48 \sqrt{2}} \csc ^{-1}\left[\sqrt{3}\right] , \nonumber \\[2mm]	
	a_8 &= \frac{265}{864} \left(\text{Li}_2\left[ \frac{3}{4}\right] + \log [4] \log \left[\frac{4}{3}\right] \right) +\frac{199393 \gamma }{138240} \nonumber \\
	& +\frac{25001310633017}{9481096396800}+\frac{4753 \pi }{13824 \sqrt{2}}+\frac{20910563 \pi }{26542080 \sqrt{3}} \nonumber \\
	& -\frac{29 \pi ^2}{288}-\left(\frac{101313035}{143327232}+\frac{804611 \pi }{442368 \sqrt{3}}\right) \log [3] \nonumber \\
	& -\left(\frac{129118553}{117573120}+\frac{804611 \pi }{294912 \sqrt{3}}\right) \log \left[\frac{4}{3}\right] - \frac{119}{288} \psi\left[\frac{5}{2}\right] \nonumber \\
	& -\frac{5}{16} \csc ^{-1}\left[\sqrt{3}\right]^2 +  \csc ^{-1}\left[\sqrt{3}\right] \bigg( \frac{823}{3072 \sqrt{2}}  \log \left[\frac{4}{3}\right] \nonumber \\
	& + \frac{5 \pi }{16} -\frac{19319 \gamma }{9216 \sqrt{2}}-\frac{5341499}{3538944 \sqrt{2}}+\frac{104075 \log [3]}{196608 \sqrt{2}} \bigg) , \nonumber \\[2mm]
	a_9 &= -\frac{8327}{138240}+\frac{804611 \pi }{884736 \sqrt{3}}-\frac{1}{96} \log \left[\frac{4}{3}\right] \nonumber \\
	& -\frac{823}{3072 \sqrt{2}}  \csc ^{-1} \left[\sqrt{3}\right] , \nonumber \\[2mm]
	a_{11} &= -\frac{5}{192} \left(\text{Li}_2\left[\frac{3}{4}\right]+\log [4] \log \left[\frac{4}{3}\right] \right)-\frac{25 \pi ^2}{192} \nonumber \\
	& -\frac{1310311 \gamma }{6635520}-\frac{10567863311827}{10113169489920} +\frac{4453 \sqrt{3} \pi }{65536} \nonumber \\
	& +\left(\frac{12616533707}{45864714240}+\frac{1674775 \pi }{7077888 \sqrt{3}}\right) \log [3] \nonumber \\
	& +\left(\frac{17720699}{46448640}+\frac{1674775 \pi }{4718592 \sqrt{3}}\right) \log \left[\frac{4}{3}\right] \nonumber \\
	& -\frac{13905571 \pi }{84934656 \sqrt{3}} -\frac{2135 \pi }{73728 \sqrt{2}}  + \frac{97}{648} \psi \left[ \frac{5}{2}\right]  \nonumber \\
	& + \frac{1}{\sqrt{2}} \bigg(\frac{605645}{18874368} -\frac{391 \gamma }{49152} - \frac{121093 \log [3]}{4194304} \nonumber \\
	& \quad -\frac{59}{4096} \log \left[\frac{4}{3}\right] \bigg) \csc ^{-1}\left[\sqrt{3}\right] , \nonumber \\[2mm]
	a_{12} &= \frac{5538437}{11612160}-\frac{1674775 \pi }{14155776 \sqrt{3}}+\frac{1}{64} \log \left[\frac{4}{3}\right] \nonumber \\
	& + \frac{59}{4096 \sqrt{2}} \csc ^{-1}\left[\sqrt{3}\right].
\label{Eq:ApproxFFnums}
\end{align}

\begin{figure}[tb]
\centering
\includegraphics[width=0.3\textwidth]{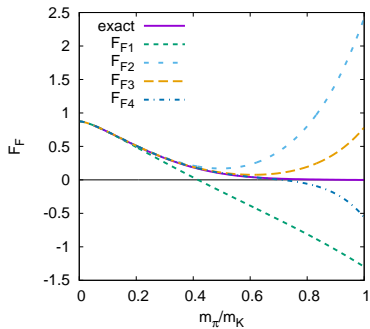}
\caption{Comparison of the exact and approximate $F_F$.}\label{Fig:FFcomp}
\end{figure}

The range of validity of Eqs.(\ref{Eq:ApproxFF})-(\ref{Eq:ApproxFFnums}) is shown in Fig.~\ref{Fig:FFcomp}, in which the exact value of $F_F$ is plotted against $x=\sqrt{\rho}$, as are the approximate $F_F$ retained up to various orders of $\rho$. The expansion up to $\mathcal{O}(\rho^4)$ approximates the exact value of $F_F$ to 1\% for $m_\pi/m_K<3$ and to 6\% for $m_\pi/m_K<0.5$. One may obtain a representation with greater accuracy by truncating the series with a larger number of terms.


For the reader to be able to verify the implementation of these expressions, we give the numerical values of $F_K/F_\pi$ coming from both exact and approximate expressions and obtained with physical values $m_\pi=0.1350$GeV, $m_K=0.4955$GeV, $F_\pi=0.0922$GeV, as well as the LEC values of the BE14 fit of \cite{Bijnens:2014lea}. We get, using Eq.(\ref{Eq:ExactFf}),
\begin{align}
	F_K/F_\pi = 1.19897,
\end{align}
and using the approximation of Eqs.(\ref{Eq:ApproxFF})-(\ref{Eq:ApproxFFnums}),
\begin{align}
	F_K/F_\pi = 1.20071.
\end{align}

\vspace*{4mm} 

{\bf Illustrative Lattice Fits}- In this section, we present an exploratory numerical study based on our analytical representation by fitting Eq.(\ref{Eq:fkfpLattice}) with the data of the lattice study \cite{Durr:2016ulb} to determine best-fit values of the NLO LEC $L^r_5$ and the NNLO LEC combinations $C^r_{14}+C^r_{15}$ and $C^r_{15}+2C^r_{17}$. We perform the fit (using \cite{James:1975dr}) on the mass sets for which $ m_\pi < 0.40$ GeV. We do the fit on the `exact' $F_F$, i.e. truncating the KdF series after $1000^2$ terms, and cross-check by fitting the exact purely numerical version of Eq.(\ref{Eq:fkfp}) with $\mathsf{CHIRON}$ \cite{Bijnens:2014gsa}. The fit on the approximate version presented in Eq.(\ref{Eq:ApproxFF}) gives compatible results. 

The uncertainties on the values of the LEC given in this section derive from the errors of the $F_K/F_\pi$ data of the lattice study, but do not take into account other uncertainties. As detailed in \cite{Durr:2016ulb}, systematic effects due to lattice artificats can arise from correlator fit time choices, lattice spacings, renormalization and finite volume corrections, among other things. When these effects are taken into account, such as by means of the results presented in \cite{Colangelo:2002hy,Colangelo:2005gd} to account for the extrapolation to infinite volume, the values of the LEC presented in this section are likely to change. However, determining the exact nature and magnitude of the change involves a detailed study that is outside the scope of this paper. Therefore, the numerical results in this section are given for an illustrative purpose only, to encourage the lattice community to undertake just such a detailed study using the NNLO analytic results presented above.

We fix the renormalization scale $\mu$ at $m_\rho = 0.77$ GeV, and use the values of the BE14 fit \cite{Bijnens:2014lea} for the other $L^r_i$. In addition we fix $F_\pi$ in the determination of $\xi_\pi$ and $\xi_K$
to 92.2~MeV and obtain:
\begin{align}
	& L^r_5 = (3.92 \pm 0.55)~10^{-4}  \nonumber \\
	& C^r_{14}+C^r_{15} = (2.59 \pm 0.63)~10^{-6} \nonumber \\
	& C^r_{15}+2C^r_{17} = (6.10 \pm 1.41)~10^{-6}. \label{Eq:LECvalues}
\end{align}

\begin{figure*}
    \centering
    \begin{minipage}{0.325\textwidth}
        \centering
        \includegraphics[width=0.9\textwidth]{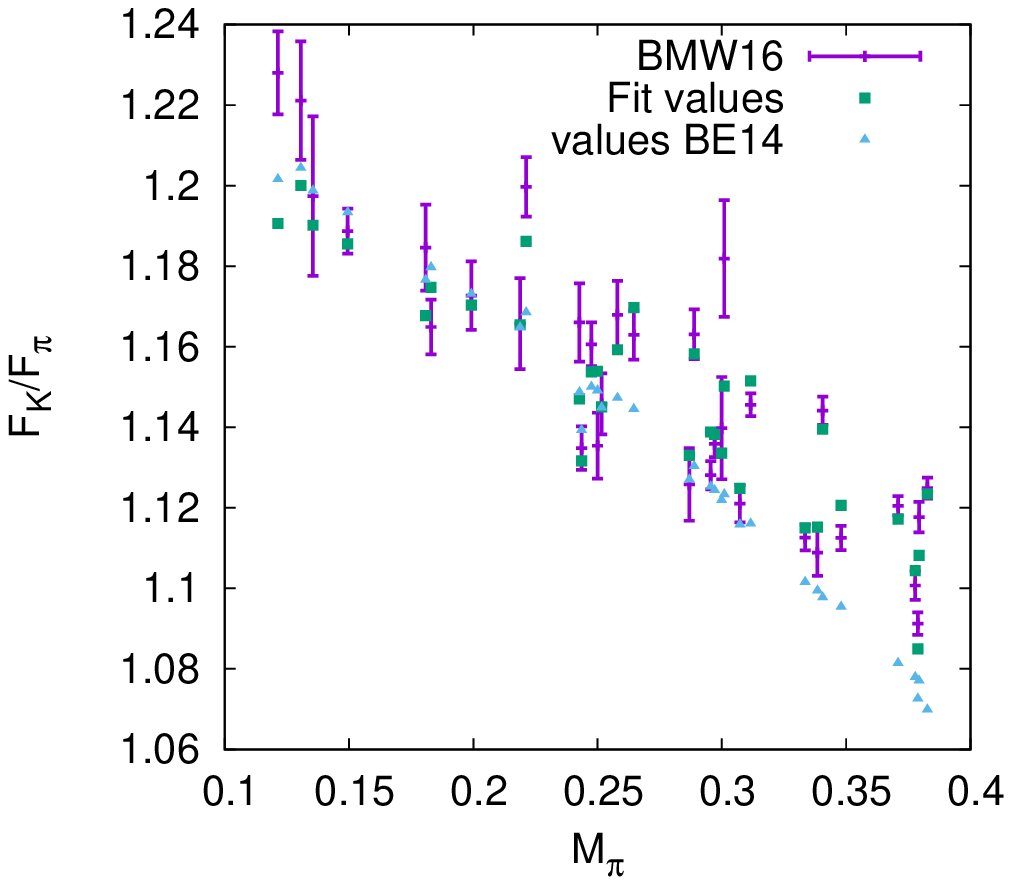} 
    \end{minipage}
    \begin{minipage}{0.325\textwidth}
        \centering
        \includegraphics[width=0.9\textwidth]{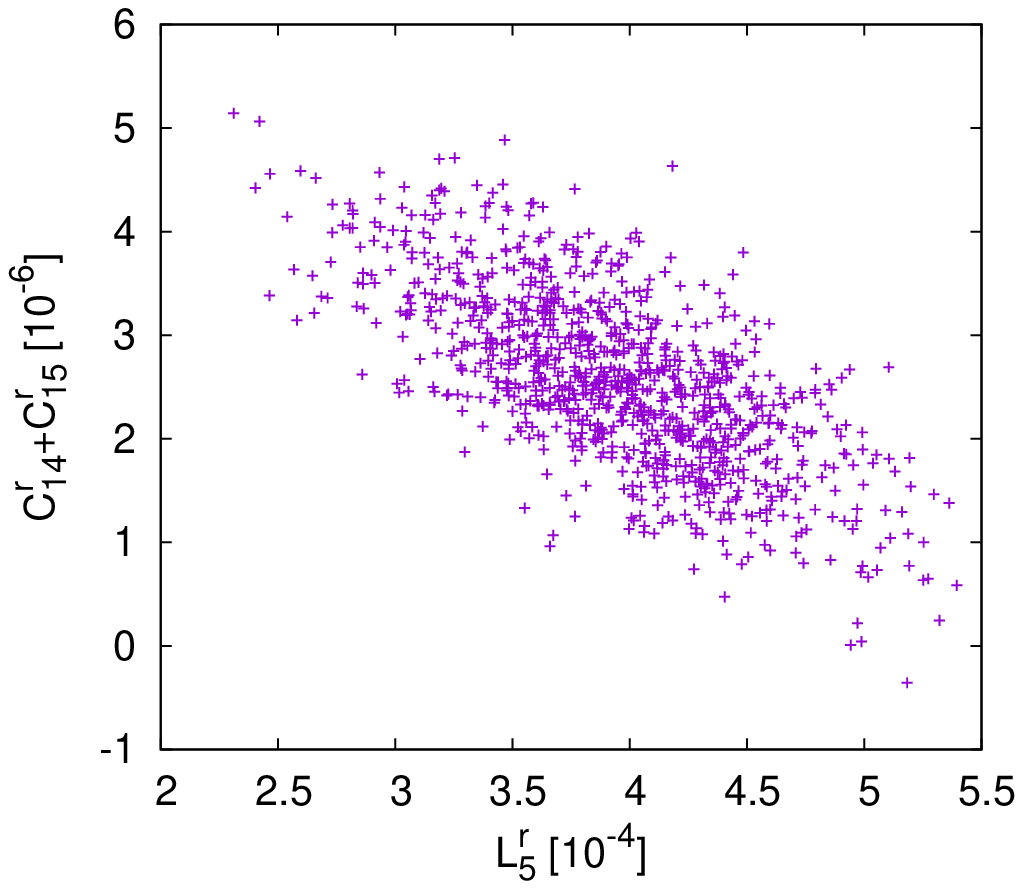}
    \end{minipage}
    \begin{minipage}{0.325\textwidth}
        \centering
        \includegraphics[width=0.9\textwidth]{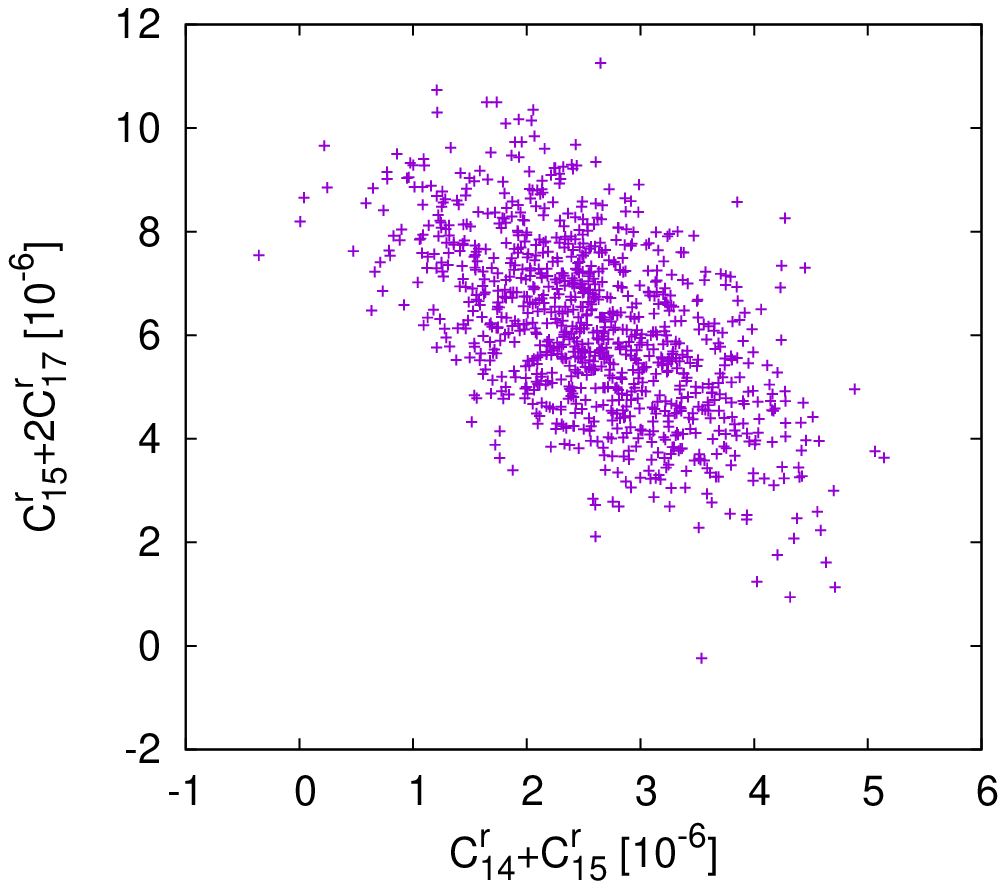}
	\end{minipage}
\caption{Left: Quality of the fit. `Values BE14' plots use the BE14 numbers for $L^r_5$, $C^r_{14}$, $C^r_{15}$ and $C^r_{17}$. Middle: Correlation of $L_5^r$ and $C^r_{14}+C^r_{15}$. Right: Correlation of $C^r_{14}+C^r_{15}$ and $C^r_{15}+2C^r_{17}$.}	\label{Fig:results}
\end{figure*}

\begin{table}
\centering
\begin{tabular}{| c | c c | }
\hline
 & $L_5$ & $C_{14}+C_{15}$  \\ 
\hline
$C_{14}+C_{15}$ & $-0.93$ & $1.00$  \\
$C_{15}+2C_{17}$ & $0.35$   & $-0.66$ \\
\hline
\end{tabular} \\
\caption{Correlation values of the fit in (\ref{Eq:LECvalues}).}
\label{Table:CorPar}
\end{table}

The correlation parameters are given in Table~\ref{Table:CorPar} and the quality of the fit is shown in Fig.~\ref{Fig:results} (Left). The correlation
is shown graphically in Fig.~\ref{Fig:results} (Middle, Right) by plotting a number of random points in a distribution given by the correlation matrix of the fit projected on the two different planes.
 
With these LEC values and the physical meson masses as inputs, we get for the value of $F_K/F_\pi$:
\begin{align}
F_K/F_\pi = 1.194,
\end{align}
which agrees well with the literature value of \cite{Bijnens:2014lea}.

The values of Eq.(\ref{Eq:LECvalues}) differ from those of the BE14 exact fit ($L_5 = 10.1 \times 10^{-4}, C_{14}+C_{15} = -4.00 \times 10^{-6} , C_{15}+2C_{17} = -5.00 \times 10^{-6}$) significantly, but are more compatible with those of \cite{Ecker:2010nc} ($L_5 = 0.76 \times 10^{-3}, C_{14}+C_{15} = 3.15 \times 10^{-6} , C_{15}+2C_{17} = 10.96 \times 10^{-6}$ in dimensionaless units) and \cite{Ecker:2013pba} ($L_5 = 0.75 \times 10^{-3}, C_{14}+C_{15} = 1.70 \times 10^{-6} , C_{15}+2C_{17} = 6.04 \times 10^{-6}$).

A similar fit, but now with $F_\pi$ also varied in $\xi_\pi,\xi_K$ requires the
use of lattices common to \cite{Durr:2016ulb} and \cite{Durr:2013goa} to obtain the values of $F_\pi$ for each lattice. This fit gives:
\begin{align}
	& L^r_5 = (0.49 \pm 1.08)~10^{-4}  \nonumber \\
	& C^r_{14}+C^r_{15} = (5.59 \pm 1.08)~10^{-6} \nonumber \\
	& C^r_{15}+2C^r_{17} = (39.7 \pm 2.10)~10^{-6}. \label{Eq:LECvalues2}
\end{align}

The change in the values above arises primarily due to the variation of $F_\pi$. Keeping $F_\pi$ fixed at 92.2~MeV but with the set of inputs used to calculate Eq.(\ref{Eq:LECvalues2}) results in changes of $\approx$ 20\%, 35\% and 10\% in the Eq.(\ref{Eq:LECvalues}) values of the $L^r_5$, $C^r_{14}+C^r_{15}$ and $C^r_{15}+2C^r_{17}$, respectively. As the difference in the inputs for Eq.(\ref{Eq:LECvalues}) and Eq.(\ref{Eq:LECvalues2}) is primarily the data from the coarsest lattices, it seems that the lattice data has a significant impact on fitting the LECs.

\vspace*{4mm}

{\bf Conclusions}- The ratio $F_K/F_\pi$ is a quantity at the heart of chiral symmetry breaking, a fundamental property of the strong interactions that is measured in ab initio calculations on the lattice. Tuning of the quark masses to physical values is now possible. Thus an analytic expansion for this quantity in masses of the quarks or the mesons is the order of the day. Using modern loop calculation techniques, we have achieved this goal. At present, two-loop precision is sufficient to fit the lattice data; this might change when the lattice precision improves in the future.
While there exist three-loop results in two-flavour ChPT \cite{Bijnens:2017wba}, in three-flavour ChPT two-loops is the state of the art, making our method and results all the more significant.

This work is a product of combining techniques developed independently in various branches of elementary particle physics and field theory, and represents an important advance on the results that appeared nearly two decades ago, when many sunsets were evaluated numerically. We hope this work will pave the way for detailed comparisons of other similar quantities with lattice simulations, and help improve our understanding of both ChPT and lattice studies.

\vspace*{4mm}

{\bf Acknowledgements}- We thank Pere Masjuan for helpful correspondance regarding the LECs. JB is supported in part by the Swedish Research Council grants contract numbers 621-2013-4287, 2015-04089 and 2016-05996 and by the European Research Council under the European Union's Horizon 2020 research and innovation programme (grant agreement No 668679). BA is partly supported by the MSIL Chair of the Division of Physical and Mathematical Sciences, Indian Institute of Science.


\begin{thebibliography}{99}


\bibitem{Durr:2016ulb}
  S.~Dürr {\it et al.},
  Phys.\ Rev.\ D {\bf 95} (2017) no.5,  054513
  doi:10.1103/PhysRevD.95.054513
  [arXiv:1601.05998 [hep-lat]].

\bibitem{Gasser:1984gg} 
  J.~Gasser and H.~Leutwyler,
  Nucl.\ Phys.\ B {\bf 250}, 465 (1985).
  doi:10.1016/0550-3213(85)90492-4

\bibitem{Amoros:1999dp}
  G.~Amoros, J.~Bijnens and P.~Talavera,
  Nucl.\ Phys.\ B {\bf 568} (2000) 319
  [hep-ph/9907264].

\bibitem{Tarasov:1997kx}
  O.~V.~Tarasov,
  Nucl.\ Phys.\ B {\bf 502} (1997) 455
  [hep-ph/9703319].
    
\bibitem{Aguilar:2008qj}
  J.~P.~Aguilar, D.~Greynat and E.~De Rafael,
  Phys.\ Rev.\ D {\bf 77} (2008) 093010
  doi:10.1103/PhysRevD.77.093010
  [arXiv:0802.2618 [hep-ph]].

\bibitem{Friot:2011ic}
  S.~Friot and D.~Greynat,
  J.\ Math.\ Phys.\  {\bf 53} (2012) 023508
  doi:10.1063/1.3679686
  [arXiv:1107.0328 [math-ph]].
    
\bibitem{Ananthanarayan:2016pos}
  B.~Ananthanarayan, J.~Bijnens, S.~Ghosh and A.~Hebbar,
  Eur.\ Phys.\ J.\ A {\bf 52} (2016) no.12,  374
  doi:10.1140/epja/i2016-16374-8
  [arXiv:1608.02386 [hep-ph]].

\bibitem{ABFG:2017}
	B.~Ananthanarayan, J.~Bijnens, S.~Friot and S.~Ghosh
	[Work in progress]

\bibitem{ABFG:2018}
	B.~Ananthanarayan, S.~Friot and S.~Ghosh
	[Work in progress]  

\bibitem{Ananthanarayan:2017yhz}
  B.~Ananthanarayan, J.~Bijnens and S.~Ghosh,
  Eur.\ Phys.\ J.\ C {\bf 77} (2017) no.7,  497
  doi:10.1140/epjc/s10052-017-5019-y
  [arXiv:1703.00141 [hep-ph]].

\bibitem{Bijnens:2014lea}
  J.~Bijnens and G.~Ecker,
  Ann.\ Rev.\ Nucl.\ Part.\ Sci.\  {\bf 64} (2014) 149
  [arXiv:1405.6488 [hep-ph]].

\bibitem{James:1975dr}
  F.~James and M.~Roos,
  Comput.\ Phys.\ Commun.\  {\bf 10} (1975) 343.
  doi:10.1016/0010-4655(75)90039-9

\bibitem{Bijnens:2014gsa}
  J.~Bijnens,
  Eur.\ Phys.\ J.\ C {\bf 75} (2015) no.1,  27
  doi:10.1140/epjc/s10052-014-3249-9
  [arXiv:1412.0887 [hep-ph]].
  http://home.thep.lu.se/~bijnens/chiron/


\bibitem{Colangelo:2002hy}
  G.~Colangelo, S.~Durr and R.~Sommer,
  Nucl.\ Phys.\ Proc.\ Suppl.\  {\bf 119} (2003) 254
  doi:10.1016/S0920-5632(03)80450-4
  [hep-lat/0209110].

\bibitem{Colangelo:2005gd}
  G.~Colangelo, S.~Durr and C.~Haefeli,
  Nucl.\ Phys.\ B {\bf 721} (2005) 136
  doi:10.1016/j.nuclphysb.2005.05.015
  [hep-lat/0503014].

\bibitem{Ecker:2010nc}
  G.~Ecker, P.~Masjuan and H.~Neufeld,
  Phys.\ Lett.\ B {\bf 692} (2010) 184
  doi:10.1016/j.physletb.2010.07.037
  [arXiv:1004.3422 [hep-ph]].

\bibitem{Ecker:2013pba}
  G.~Ecker, P.~Masjuan and H.~Neufeld,
  Eur.\ Phys.\ J.\ C {\bf 74} (2014) no.2,  2748
  doi:10.1140/epjc/s10052-014-2748-z
  [arXiv:1310.8452 [hep-ph]].

\bibitem{Durr:2013goa} 
  S.~Dürr {\it et al.} [Budapest-Marseille-Wuppertal Collaboration],
  Phys.\ Rev.\ D {\bf 90}, no. 11, 114504 (2014)
  doi:10.1103/PhysRevD.90.114504
  [arXiv:1310.3626 [hep-lat]].

  
\bibitem{Bijnens:2017wba}
  J.~Bijnens and N.~H.~Truedsson,
  JHEP {\bf 1711} (2017) 181
  doi:10.1007/JHEP11(2017)181
  [arXiv:1710.01901 [hep-ph]].


\end{thebibliography}
\end{document}